\documentclass{eas}
\usepackage{graphicx}
\newcommand{\kms}{\,km\,s$^{-1}$}
\newcommand{\myr}{\,$M_{\odot}\,{\rm yr}^{-1}$}
\newcommand{\ro}{\,$R_{\odot}$}

\newcommand{\lo}{\,$L_{\odot}$}
\begin{document}

\title{Evolved stars as donors in symbiotic binaries}
\author{A. Skopal}\address{Astronomical Institute of the Slovak Academy 
                           of Sciences, Tatransk\'a Lomnica, Slovakia}
\author{M. Seker\'a\v{s}}\sameaddress{1}
\author{N. Shagatova}\sameaddress{1}
\runningtitle{Skopal \etal: Evolved donors in symbiotic stars}
\thanks{Supported by a grant of the Slovak Academy of Sciences, 
        VEGA No.~2/0002/13.}
\begin{abstract}
This contribution is focused on the role of cool giants in 
symbiotic binaries. Especially, we pay attention to their 
mass-loss rates and the wind mass-transfer onto their compact 
accretors. 

\end{abstract}
\maketitle
\section{Introduction}
Symbiotic stars (SSs) are the widest interacting binaries, whose 
orbital periods run from hundreds of days to hundreds of years. 
They consist of an evolved red giant (RG) as the donor star and 
a white dwarf (WD) accreting from the giant's wind. The accretion 
process heats up the WD to $>10^5$\,K and increases its 
luminosity to $\sim 10^{2}-10^{4}$\lo, which by return ionizes 
a fraction of the neutral wind from the giant giving rise to 
the nebular radiation (Seaquist \etal\ \cite{stb}, hereafter STB). 

In most cases, the observed large energetic output is believed
to be caused by stable nuclear hydrogen burning on the WD  
surface, which requires accretion onto a low mass WD at 
$10^{-8} - 10^{-7}$\myr\ (e.g. Shen \& Bildsten \cite{shen+07}). 
However, such high accretion rates cannot be achieved by 
a standard Bondi-Hoyle wind accretion, because of its small 
efficiency (Bondi \& Hoyle \cite{b+h44}), and the mass-loss 
rates of $\approx 10^{-7}$\myr\ from RGs in S-type SSs 
(e.g. Seaquist \etal\ \cite{s+93}). This problem was pointed already 
by Kenyon \& Gallagher (\cite{ken+gall83}). 

Accordingly, we introduce principle of methods to estimate 
the mas-loss rate from RGs in SSs (Sect.~2), and possible 
solutions of the required high mass-transfer ratio (Sect.~3). 
\begin{figure}
 \includegraphics[width=4.2cm,angle=-90]{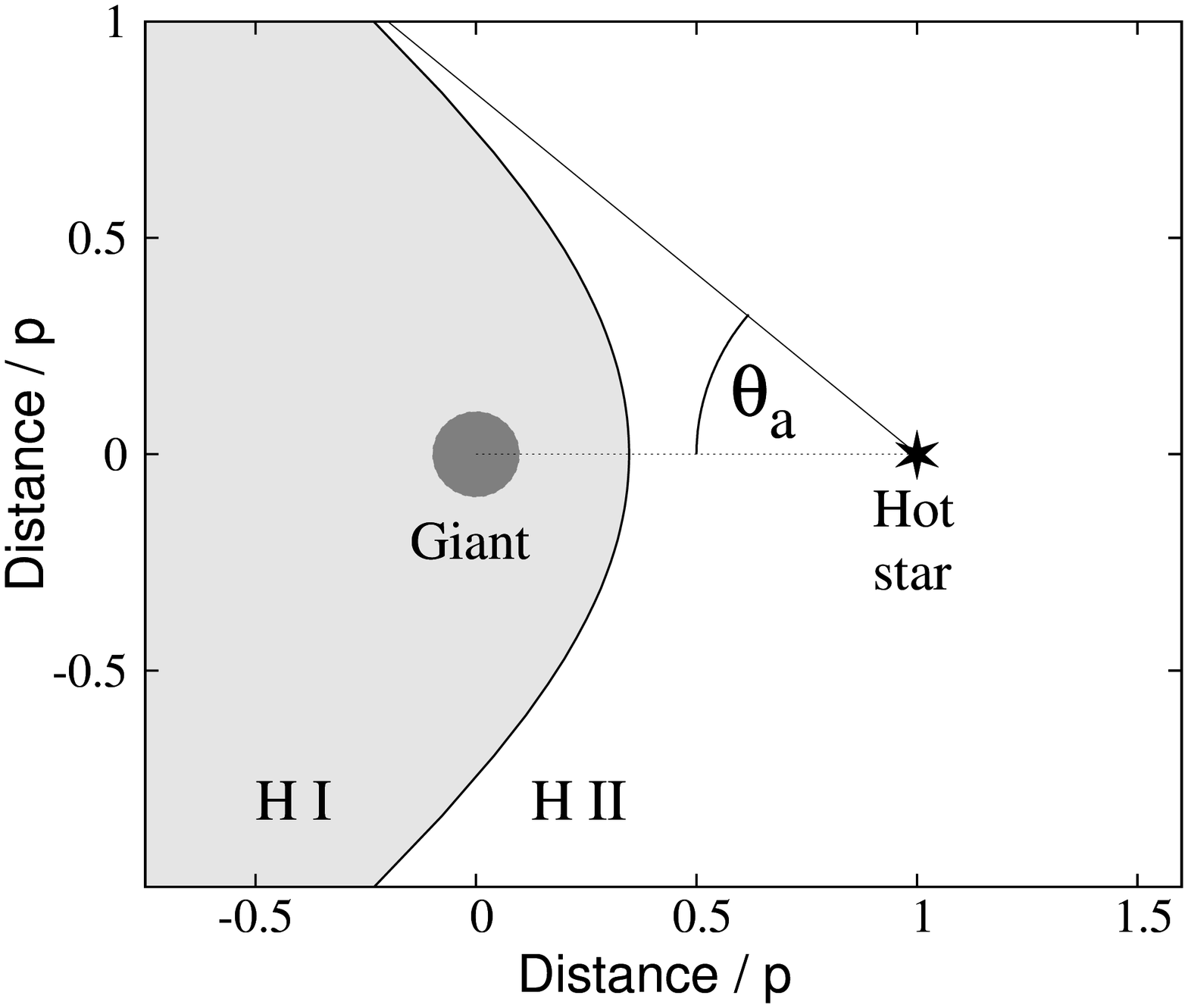}
 \qquad
 \includegraphics[width=4.1cm,angle=-90]{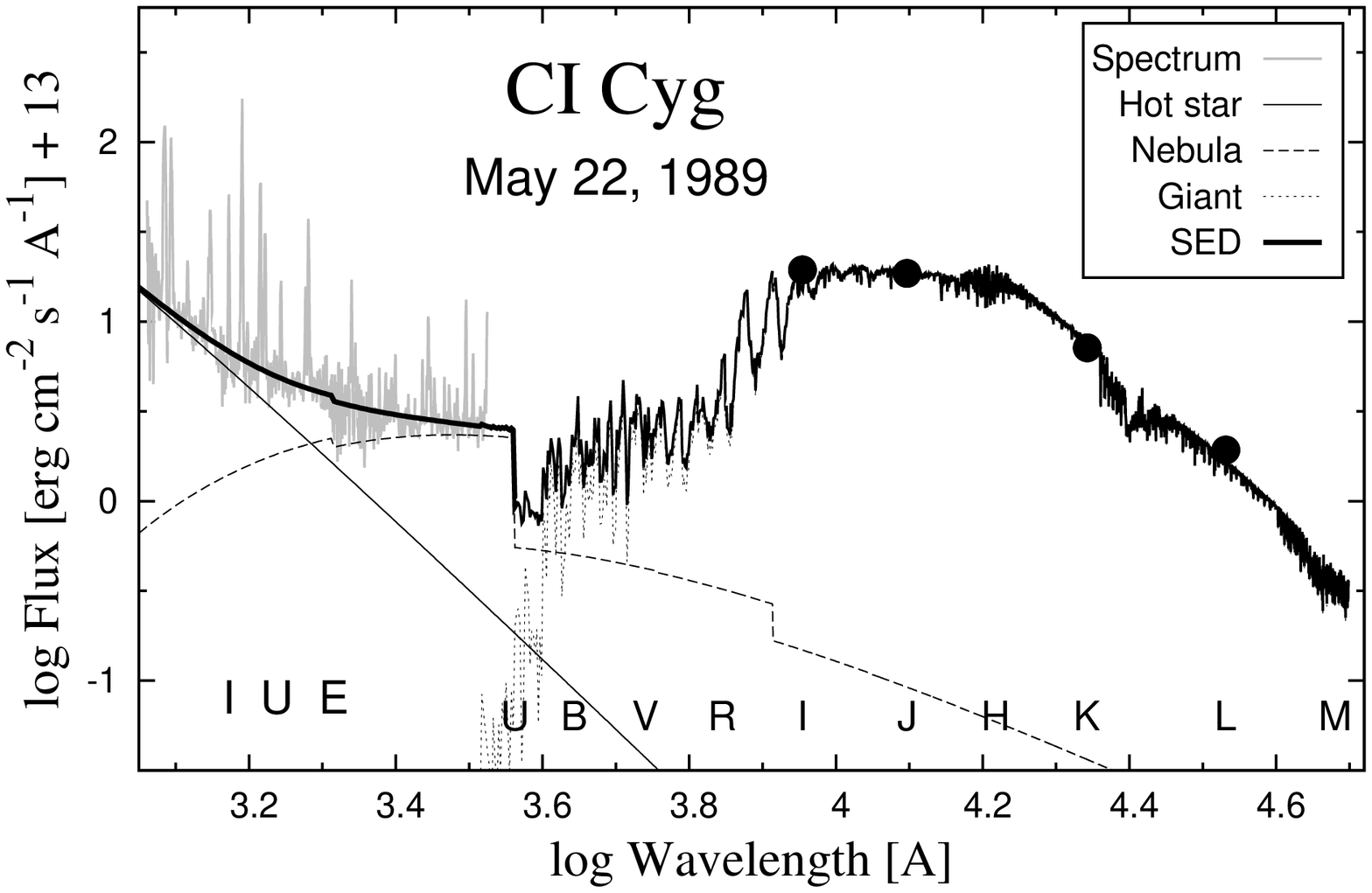}
\caption{The STB ionization structure (left) and 
corresponding SED for CI~Cyg (right). Extension 
of the neutral zone is determined by the asymptotic 
angle $\theta_{\rm a}$.} 
\end{figure}

\section{Mass-loss rates from RGs in symbiotic binaries}
Mass-loss rates from giants in symbiotic binaries can be 
determined in the context of the STB model by measuring 
the wind nebular emission during quiescent phase. 
The extent of the ionized RG wind is given by a parametric 
equation, 
\begin{equation}
   f(u,\vartheta) - X = 0,
\label{XH+=f}
\end{equation}
the solution of which defines the H\,{\small I}/H\,{\small II} 
boundary at the orbital plane determined by polar coordinates 
($u,\vartheta$) centered at the hot star for a stationary binary. 
The parameter $X$ is given by the binary properties 
and the wind mas-loss rate, $\dot M_{\rm RG}$ (see STB). 
Figure~1 depicts the STB ionization structure and the 
corresponding typical UV/near-IR SED of SSs. 
Measuring the radio emission at 3.6\,cm for a sample of 99 
SSs, Seaquist \etal\ (\cite{s+93}) derived 
$\dot M_{\rm RG} \approx 10^{-7}$\myr. 

In the UV/optical/near-IR, the nebular emission can be obtained 
by modelling the SED (see example in Fig.~1, right). 
On the other hand, one can calculate the nebular emission 
by integrating contributions throughout the volume of the fully 
ionized zone (see Fig.~1, left), which depends on parameters 
of the giant's wind. Comparing the observed and calculated 
continuum nebular emission, we get $\dot M_{\rm RG}$ for 
a given wind terminal velocity. In this way, Skopal (\cite{sk05}) 
determined $\dot M_{\rm RG}$ = a few $\times 10^{-7}$\myr\ 
for giants in 15 well observed S-type SSs. 

$\dot M_{\rm RG}$ can also be determined by probing directly 
the neutral fraction of the RG wind in SSs. 
Here, Raman scattering of the far-UV line photons on atomic 
hydrogen in the wind is investigated. The key parameter is 
the efficiency of this process, defined as the ratio between 
the Raman scattered and the original line photons. 
The Raman scattering efficiency defines the so-called `covering 
factor' $C_{\rm S}$, which represents a fraction of the sky 
`seen' from the emission zone located predominately near the 
hot component, which is covered by the Raman scattering region. 
Assuming the STB geometry for the neutral zone, we can express 
$C_{\rm S}$ via a solid angle $\Omega$, under which the initial 
line photons can `see' the scattering region, 
%
\begin{equation}
  C_{\rm S} = \frac{\Omega}{4\pi} = 
              \frac{1-\cos\theta_{\rm R}}{2}, 
\label{eq:omega}              
\end{equation}
where $\theta_{\rm R}$ is the opening angle of the Raman scattering 
region. If the STB neutral zone is optically thick for Raman 
scattering, then $\theta_{\rm R} \sim \theta_{\rm a}$, which 
determines unambiguously the parameter $X$ (see Fig.~1, 
left). Otherwise, one has to reconstruct $\theta_{\rm a}$ 
from $\theta_{\rm R}$ taking into account optically thick 
conditions for the investigated Raman scattering conversion. 
Finally, having the parameter $X$ and the fundamental 
parameters of the hot component, one can derive $\dot M_{\rm RG}$. 
Using Raman He\,{\small II}\,$\lambda 1025\rightarrow \lambda 6545$ 
conversion we determined $\dot M_{\rm RG} = 2-3 \times 10^{-6}$\myr\ 
for the mira-type variable in V1016~Cyg 
(see Seker\'a\v{s} \& Skopal, this proceedings).
\begin{figure}
\begin{center}
 \includegraphics[width=4cm,angle=0]{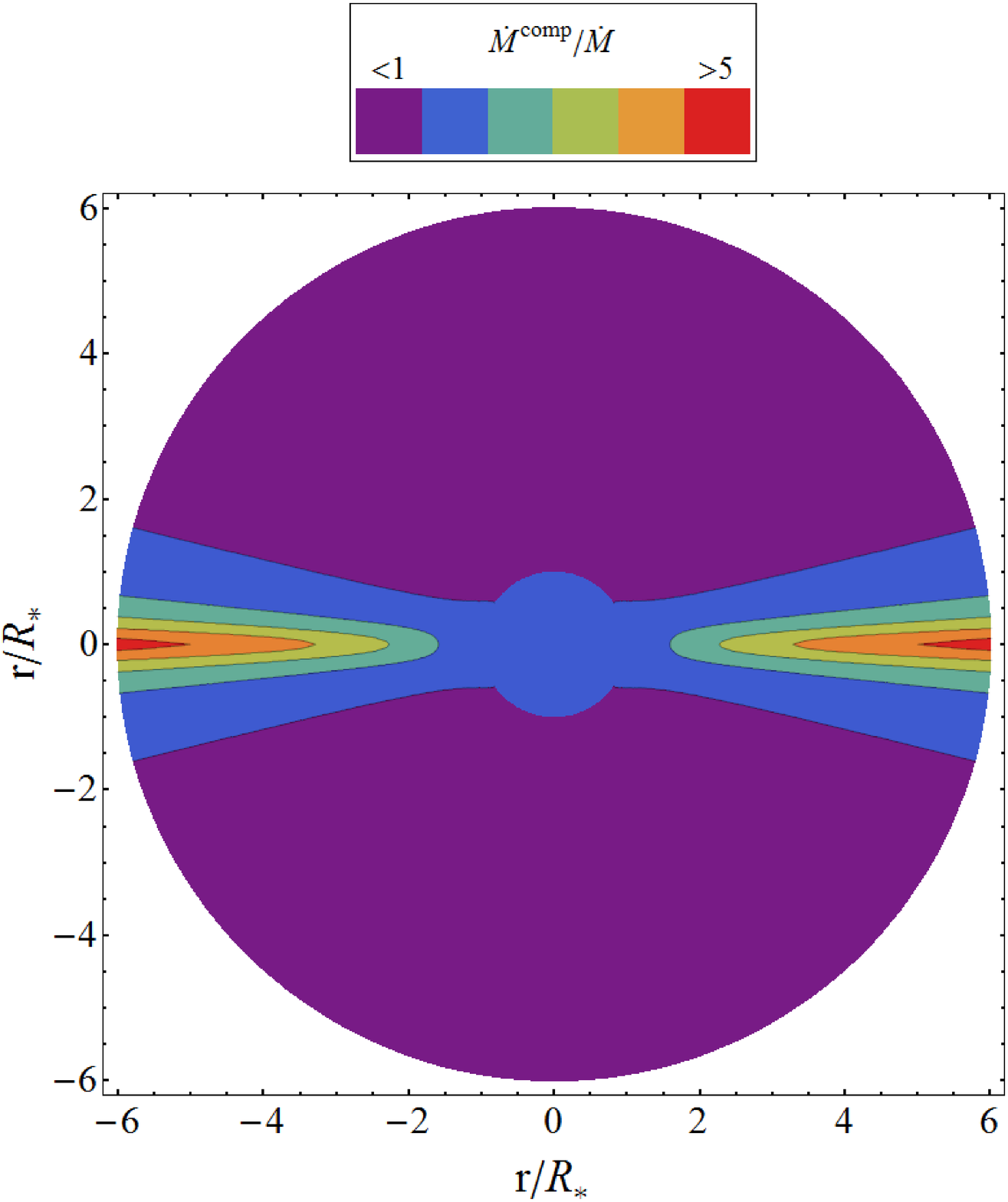}
 \qquad
 \includegraphics[width=4cm,angle=0]{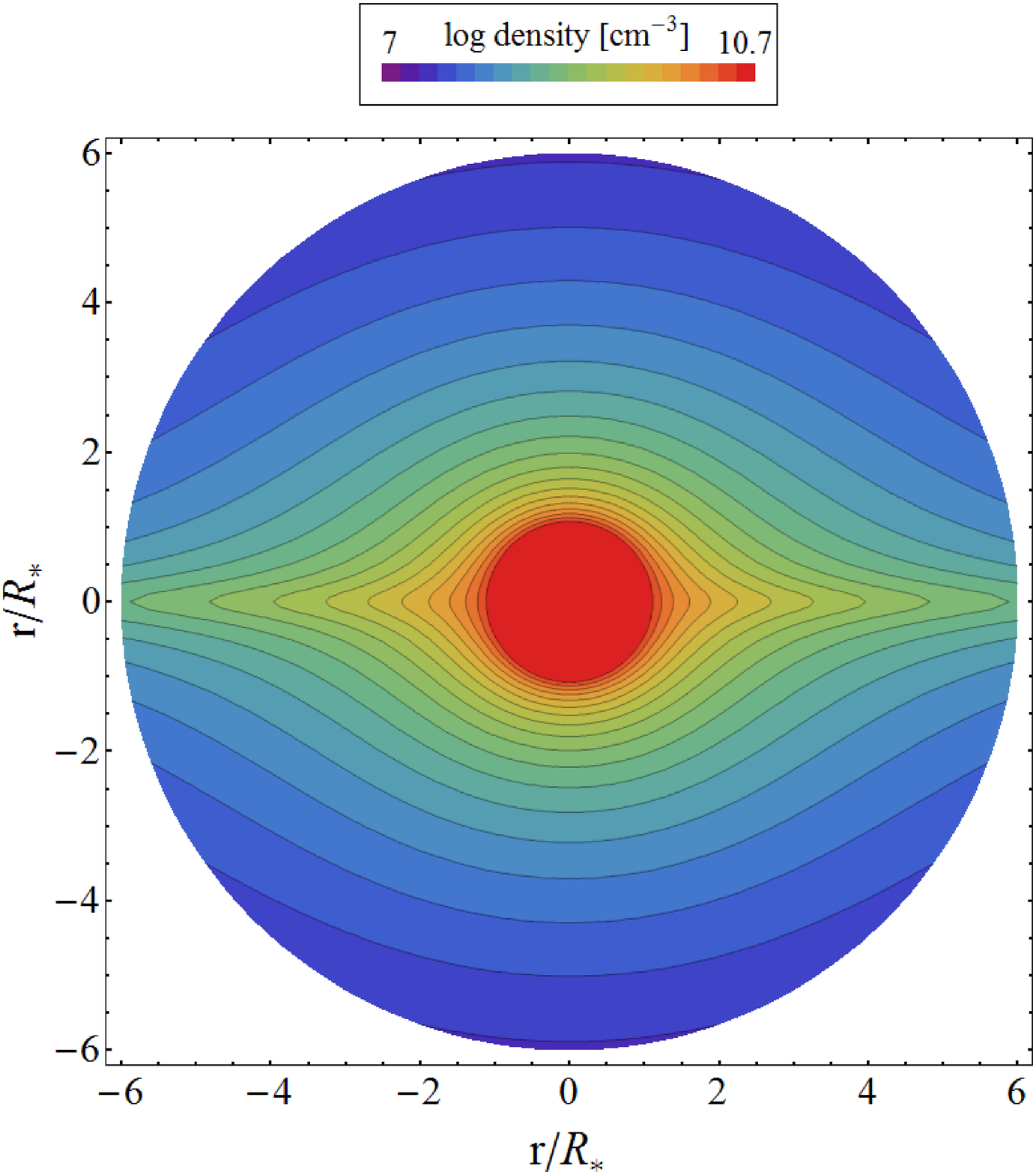}
\end{center}
\vspace*{-5mm}
\caption{Left: Compression of the wind to the equatorial 
plane relative to the spherically symmetric case. 
Right: Corresponding density distribution. Both calculated 
according to the wind compression model for giant's rotation 
of 6\kms, wind terminal velocity of 20\kms, 
$\dot M_{\rm RG} = 10^{-7}$\myr\ and $R_{\rm RG} = 100$\ro\
(see Skopal \& Carikov\'a \cite{sc15}). 
        } 
\end{figure}

\section{On the mass-transfer ratio in symbiotic binaries}

The long-standing problem of the large energetic output from
the majority of hot components in symbiotic binaries and their 
deficient fueling by the RG's wind in the canonical Bondi-Hoyle 
picture was recently approached in two ways. 

An efficient mass-transfer mode was suggested for Mira-type 
interacting binaries (i.e. being in the effect for D-type SSs) 
by Mohamed \& Podsiadlowski (\cite{mp12}). 
In this case, a slow and dense wind from an evolved AGB star 
is filling the Roche lobe ($v_{\rm wind} < v_{\rm escape}$) 
instead the star itself, and thus can be transferred very 
effectively into the potential of the companion via 
the $L_1$ point. This mass transfer mode is called 
{\em wind Roche-lobe overflow} (WRLOF). 

In the case of S-type SSs, whose donors are normal 
RGs, an effective wind mass transfer can be caused by their 
rotation. Recently, Skopal \& Carikov\'a (\cite{sc15}) applied 
the wind compression disk model (WCD) of 
Bjorkman \& Cassinelli (\cite{bjorkcass93}) to slowly 
rotating giants in S-type SSs, and found that their wind can 
be focused at the equatorial plane with a factor of 5--10 
relative to the spherically symmetric wind (Fig.~2). This 
suggests a relevant increase of the accretion rate onto the WD. 
Investigating the hydrogen column densities, we obtained from 
the spectra of eclipsing S-type SSs, suggests that the wind 
from their RGs is really enhanced at the orbital plane 
(see Shagatova \& Skopal, this proceedings). 

\section{Conclusions}

Mass-loss rates from RGs in SSs are in the order of $10^{-7}$ 
and $10^{-6}$\myr\ for S-type and D-type systems. The high 
luminosities of their accretors require an efficient wind 
mass-transfer mode. For D-type SSs, the WRLOF mode 
can be considered, while for S-type SSs, the WCD model 
can be in the effect. 

%

\noindent {\bf Discussion}\\
J. Mikolajewska: 
STB model predicts the relation between turn-over frequency 
and orbital separation. The symbiotic star CI~Cyg with very 
good radio spectrum (simultaneous VLA and SCUBA) shows that 
this model cannot be applied for the S-type symbiotic stars. 
Can you comment on this? \\
A. Skopal: 
STB model describes the simplest ionization structure of 
SSs as given by their nature. Its applicability is, 
however, restricted by its simplicity (e.g. stationary 
binary). As concerns to CI~Cyg, its UV/optical/near-IR SED 
is consistent with the STB model (see Fig.~1). \\
J. Mikolajewska: 
SY Mus has strong ellipsoidal variability which requires much 
larger radius than your model suggests. \\
A. Skopal: 
The RG's radius in SY~Mus of 86\ro was derived from its 
rotation velocity assuming its co-rotation with the orbit. 
Similar discrepancy is indicated by other SSs, for 
example, FG~Ser. \\
A. Lobel: 
RR~Tel is an important symbiotic object, because of its very 
rich UV forest of emission lines from high ionic transitions. 
It is often used for determinations of fundamental atomic 
data of coronal emission lines that cannot be determined from 
laboratory measurements. Do you plan to measure the mass-loss 
rate of RR Tel as well using your method? \\
A. Skopal: 
Yes, of course. The only obstacle is to obtain its spectrum 
($\delta \sim -56^{\circ}$). \\
A. Miroshnichenko: 
The wind compression model was rejected for Be stars. The disk 
forms only if the radiation pressure produces only radial forces. 
Was applicability of the WCD model tested for symbiotic stars? \\
A. Skopal: 
Yes, we tested the applicability of the WCD model to SSs in 
our previous papers. For example, the rotational flattening for 
parameters of a typical RG in SSs is only 0.975. 
\end{document}